\documentclass[screen,10pt]{acmart}
\usepackage{enumitem}
\newcommand{\beq}{\begin{equation} \begin{array}{l} }
\newcommand{\eeq}{\end{array}\end{equation}}
\newcommand{\beqq}{\begin{equation*} \begin{array}{l} }
\newcommand{\eeqq}{\end{array}\end{equation*}}
\newcommand{\ess}{\epsilon}
\newcommand{\xy}{\cdot}
\newcommand{\Tau}{\mathit{U}}

\newcommand{\concat}{\mathop{\mathit{concat}}}

\newcommand{\txd}{\mathit{Sent}}

\newcommand{\rxd}{\mathit{Received}}

\newcommand{\Msgs}{\mathit{Messages}}

\newtheorem{lemma}{Lemma}
\newtheorem{definition}{Definition}[section]
\newtheorem{theorem}{Theorem}
\newcommand{\set}[1]{\{#1\}}

\setcopyright{acmlicensed}
\acmJournal{FAC}
\acmYear{2023}
\acmVolume{1}
\acmNumber{1}
\acmArticle{1}
\acmMonth{1}
\acmDOI{10.1145/3633786}

\begin{document}

%%
%% The ''title'' command has an optional parameter,
%% allowing the author to define a ''short title'' to be used in page headers.
\title{State machines for large scale computer software and systems}

\orcid{0000-0001-5085-9794}

%%
%% The ''author'' command and its associated commands are used to define
%% the authors and their affiliations.
%% Of note is the shared affiliation of the first two authors, and the
%% ''authornote'' and ''authornotemark'' commands
%% used to denote shared contribution to the research.
\author{Victor Yodaiken}
\email{vy@e27182.com}
\affiliation{%
  \institution{Independent Researcher}
  \streetaddress{2718 Creeks Edge Parkway}
  \city{Austin}
  \state{Texas}
  \country{USA}
  \postcode{78733}
}

%% The abstract is a short summary of the work to be presented in the
%% article.
\begin{abstract}
The behavior and architecture of large scale discrete state systems found in computer software and hardware
can be specified and analyzed using a particular class of primitive recursive 
functions.  This paper begins with an illustration of the 
utility of the method via a number of small examples and then via  longer 
specification and verification of the ''Paxos'' distributed consensus algorithm\cite{paxosmadesimple}. The 
``sequence maps''
are then shown to  provide  an alternative representation of deterministic state machines and products of state machines. 

Distributed and composite systems, parallel and concurrent computation, and real-time behavior can all be specified naturally with these methods - which 
require neither extensions to the classical state machine model nor any axiomatic methods or other techniques from formal logic or other
	foundational methods. 
Compared to 
state diagrams  or tables or the standard set-tuple-transition-maps, sequence maps are more concise and better suited to describing the behavior
and compositional architecture of  computer systems. 
Staying strictly within the boundaries of classical deterministic state machines 
anchors the methods to the algebraic structures of automata and makes the specifications faithful to engineering practice.
\end{abstract}




\maketitle

\section{Introduction}
This paper has three objectives:
\begin{itemize}
\item To introduce a method involving certain recursive functions 
for  specifying and verifying large scale discrete state systems. 
\item To illustrate the application of the method to analysis of problems in computer systems design including distributed consensus network algorithms and real-time.
\item To precisely define the class of recursive functions of interest and show how they relate to  classical state machines, automata products, and algebraic automata theory.
\end{itemize}

The problem addressed here  is not how to validate code,
but how to  understand and validate \emph{designs} prior to implementation and test. 
The motivation comes out of the author's experience 
designing, coding, and managing design and development projects
for operating systems, real-time, and other
 ``systems''  \cite{auragen,usenixyb,idle,yodaikenclock} without any satisfactory method for even specifying design goals\footnote{Such a judgment is necessarily at least somewhat subjective. See 
 section \ref{sec:related} for more discussion.}.  The solution proposed here is based on Moore type automata - deterministic state machines with an output associated with each state \cite{Hopcroft,Moore} . 
Since the 1960s, it has been known that digital systems can be faithfully described by 
state machines and interconnected digital systems by state machine products, but that the standard methods for representing 
state machines do not scale well. This paper shows how state machines can be defined via a certain class of recursive functions so that for a finite sequence of events, \(w\), \(f_M(w)\) will be the output of state machine \(M\) in the state \(M\) reaches by following \(w\) from its initial state (see figure \ref{fig:map}).
\begin{figure}[ht]
\[\begin{array}{rcl} \mbox{Input sequence }w \rightarrow& \colorbox{blue}{\textcolor{white}{Moore Machine \(M\)}}&\rightarrow \  \mbox{Output } = f_M(w)\\
 \end{array}\]
 \caption{Sequence maps representing state machines}
 \label{fig:map}
\end{figure}

 Function composition can be used to build up  more complex state system descriptions
 from simpler ones and also, perhaps unexpectedly, to describe  interconnected composite systems with components that change state in parallel. 
This approach permits concise but exact definitions of state machines and abstract
 properties of state machines that are completely impractical
 for state diagrams or transition maps where state sets need to be enumerated.
 In particular, state machines constructed with general automata products
 \cite{Hartmanis,Gecseg},  in which arbitrary communication can be specified via ``feedback'', can be defined
 in terms of function composition (see figure \ref{fig:general1}). 
\begin{figure} 
\[\mathit{input\ } w \rightarrow
\fbox{
\(\begin{array}{l}
\rightarrow  u_1 \rightarrow \colorbox{blue}{\textcolor{white}{\(M_1\)}}\rightarrow x_1 \\
\rightarrow  u_2\rightarrow \colorbox{blue}{\textcolor{white}{\(M_2\)}}\rightarrow x_2 \\
\dots\\
\rightarrow  u_n \rightarrow \colorbox{blue}{\textcolor{white}{\(M_n\)}}\rightarrow x_n \\
\\
\quad\textcolor{red}{\Uparrow} \quad\quad\textcolor{red}{feedback \Downarrow}\\
\quad	\begin{array}{c}\textcolor{red}{\Leftarrow  (x_1,\dots x_n)\Leftarrow}\end{array}
\end{array}\)
} \rightarrow \mbox{Output }(x_1,\dots x_n) 
\]
 \caption{The general product}
 \label{fig:general1}
\end{figure}

 In the hopes that the basic ideas are reasonably intuitive,  section \ref{sec:small} is an attempt to explain the method
 via examples
 and section \ref{sec:big} specifies and validates the notoriously slippery Paxos
 distributed consensus algorithm\cite{paxosmadesimple} as an indication that the methods can be applied to substantial problems. 
Section \ref{sec:math} defines the class of primitive recursive sequence
functions and shows precisely how they relate to Moore type automata
and Moore machine products.
Readers who are skeptical of the appeals
to intuition in sections \ref{sec:small} and \ref{sec:big} might want to read section \ref{sec:math} first. 
Section \ref{sec:related} discusses motivation, background, and differences from related work in both formal methods and
algebraic automata theory.

\section{\label{sec:small} Introduction to applications and method}
Maps on finite sequences\footnote{All sequences of events in this paper are finite.}
can be concisely defined with ``primitive recursion on words''
\cite{PeterComputer}.
Let \(\ess\) be the empty sequence and \(w\xy a\) be the sequence obtained by appending event \(a\) to sequence 
\(w\) on the right. Then \(f(\ess)\) is the
output in the initial state and given some  constant \(c\) and 
map \(g\)
\[f(\ess)=c,\mbox{ and } f(w\xy a)= g(f(w),a)\]
defines \(f\) in every reachable state\footnote{ More precisely, suppose we have a set \(A\) of events,
a set \(Y\) of outputs, some constant \(c\in Y\), and \(g:Y\times A\to Y\), then let \(f(\ess)=c\) and for all \(a\in A\) and
\(w\in A^ *\), \(f(w\xy a)=g(f(w),a)\).}.   To illustrate:
\[ Counter(\ess) = 0\mbox{ and } Counter(w\xy a)= Counter(w)+1\bmod k\]
counts events \(\bmod\ k\) for positive integer \(k\). 
\(Counter\) is really a family of finite state systems with \(k\) as a parameter.
The equations \(\Tau(\ess)=0\) and \(\Tau(w\xy a) = \Tau(w)+1\) define an
unbounded counter\footnote{A wary reader might note here that despite all the 
protestations about classical automata theory, \(\Tau\) does not describe
a \emph{finite} state Moore machine. More on this in sections \ref{sec:math} and
\ref{sec:related} , but while finiteness
is a critical and necessary property of systems, sometimes we don't know or care about the bound and sometimes it's useful to have possibly unbounded imaginary
systems to help measure or constrain behavior.}.  

Partially specified sequence maps are often useful. 
Suppose for some constant  \(0\leq c< k\) and mod \(k\) counter, \(Counter\),
\[ |(C(w)- Counter(w)) | \leq c.\]
Then \(C(w)\) is a counter with error of \(c\) or less.
Any sequence map that is a solution to this inequality has a useful property of counting within the error ranges,
but we are not required to specify exactly what the value is in any particular state.
Sequence maps specify deterministic systems but systems of interest are often not completely specified. Rather  
than positing some irreducible ``non-determinism'', we can think of these maps as solutions to constraints which may have many different solutions. 

Ordinary function composition can
modify the output. For some constant \(c\) and the  mod \(k\) counter,  \(Down(w) = k - Counter(w)\) is a down counter to
\(1\). 

Or  suppose \(C_1\) is a mod \(k_1\) counter and \(C_2\) is a mod \(k_2\) counter and 
\[F(w) = \begin{cases} 0&\mbox{if }C_1(w)+C_2(w)=0\\
         1&\mbox{otherwise}
         \end{cases}\]
Then \(F(w)=0\) only when the number of events is  divisible by both  \(k_1\) and  \(k_2\).
Suppose the alphabet of events is \(\set{0,1}\)  and we want to count consecutive 1's with 
``saturation'' counter 
\[ RCounter(\ess)=0,\quad RCounter(w\xy a) = \begin{cases} 
	0&\mbox{if }a= 0;\\
                                            RCounter(w)+1&\mbox{if }a=1\mbox{ and }RCounter(w)<k\\
                                            Rcounter(w)&\mbox{otherwise.}\end{cases}\]
\(RCounter(w)=k\) if the last \(k\) or more events have been equal to\(1\).

\subsection{Shift register and connection maps}
In addition to modifying the output, with composition, \(h(f(w))\), it is 
possible to modify the input by making the event sequence a dependent
variable where \(u(w)\) translates \(w\) to a different event sequence and \(f(u(w))\) is the output of \(f\)
in the state determined by \(u(w)\). This is how networks and 
any other interconnected systems in the examples are
specified. To get an intuition of how connections work, consider an elementary example: an idealized shift register
constructed by connecting storage cells. 
A storage cell is defined trivially by 
\beq\label{eq:cell} Cell(w\xy a) = a\eeq
to just 
remember the most recent event. The cells are connected in to each other as in figure \ref{fig:shift}
by connection maps \(u_i\) defined below so that \(Cell(u_i(w))\) is the value stored in the \(i^{th}\) cell.
\begin{figure}[ht]
	\[\begin{array}{ccc} w \rightarrow&\colorbox{blue}{\textcolor{white}{\(\begin{array}{|c|c|c|c|c|}\hline
	\mathit{Cell}_1 & \mathit{Cell}_2& \mathit{Cell}_3& \dots &\mathit{Cell}_k\\ \hline \end{array}\)}}&\rightarrow 
 \end{array}\]
 \caption{Shift register }
 \label{fig:shift}
\end{figure}

Define 
\[ u_i(\ess)=\ess,\quad u_i(w\xy a) = \begin{cases}
	u_i(w)\xy a&\mbox{if }i=1\\
	u_i(w)\xy Cell(u_{i-1}(w))&\mbox{if }1< i \leq k
\end{cases}\]

Consider \(Cell(u_1(w))\). The initial value is unspecified so \(Cell(u_1(\ess))=Cell(\ess)\) is undefined. 
Suppose the first event is \(1\), then \(Cell(u_1(\ess\xy 1))=1\) and \(Cell(u_2(\ess\xy 1\xy 2)) = 1\) and \(Cell(u_1(\ess\xy 1\xy 2)) = 2\).

The connection maps follow a standard scheme \(u_d(\ess)=\ess\) so the components all start in their
initial states, and either \(u_d(w\xy a) = u_d(w)\xy b\) for some event \(b\) in the event alphabet of the component, 
or \(u_d(w\xy a) = u_d(w)\) to leave the component state unchanged 
\footnote{ Where useful, the components can be defined to advance by multiple
steps on a single event of the enclosing system. See section \ref{sec:generalf}}. 

There's a definition of a memory array in appendix \ref{appendix:memory}.

\subsection{A controller for a physical plant\label{sec:controller}}
Consider a control system where inputs are samples of some signal. 
A simple controller that produces a control signal, might have a fixed set point \(\kappa_0\) and an error:
\[E(\ess)=0\mbox{ and } E(w\xy y) = \kappa_0 - y\]
where \(y\) is a signal sample. 
The change in the last error signal (the ``derivative'') is
\[D(\ess)=0\mbox{ and }\quad D(w\xy y) = E(w)-(\kappa_0 -y) .\]
The sum of all errors is:
\[S(\ess) = 0\mbox{ and }\quad S(w\xy y ) = S(w)+ (\kappa_0 -y)\]
although we would probably want, at some point of development, to force a bound on this sum or to show that it is bounded by some external condition such as an oscillation 
around the x-axis. 
Then the control signal might be  calculated using three parameters as follows.
\beq \label{eq:control} Control(w) = \kappa_1 E(w)+ \kappa_2 D(w) + \kappa_3 S(w).\eeq

Let
\[v(\ess)=\ess,\ v(w\xy a) = \begin{cases}v(w)\xy 1 &\mbox{if }E(w) > \kappa_4\\ v(w)\xy 0 &\mbox{otherwise}\end{cases}\]
	Here '1' and '0' are treated as event symbols.
\(RCounter(v(w))=k\) if and only if the error has been greater than \(\kappa_4\) for at least the
last \(k\) time units. 

If there is a physics model of the plant, then \(\Tau(w)\) should approximate  the real valued time variable. For example, we could require that if \(t\) is a real-valued time variable in seconds that
\(\Tau(w) \approx t\), perhaps so that it is within 10 picoseconds or whatever level of precision is appropriate.

Imagine there are \(n\) control systems embedded within a system 
so that events are vectors \(\vec{x} = (x_1,\dots x_n)\) where \(\vec{x}_i = x_i\). A map \(u_i(w)\) can
isolate the inputs on line  \(i\) as follows:
\[u_i(\ess) = \ess,\quad u_i(w\xy \vec{x}) =  u_i(w)\xy \vec{x}_i\]
Then \(Control_i(w) = Control(u_i(w))\) is a controller that operates as 
specified  by equation \ref{eq:control}  but on the indicated signal.  
This is the simplest parallel system- with no interaction between components (see figure \ref{fig:directc}).

\begin{figure}[ht]
\[\mbox{input sequence }w\xy \vec{x} \rightarrow 
\begin{array}{lcl}
\fbox{\(\begin{array}{l}
w \xy x_1\rightarrow \colorbox{blue}{\textcolor{white}{\(Control_1\)}}\rightarrow y_1 \\
w\xy x_2\rightarrow \colorbox{blue}{\textcolor{white}{\(Control_2\)}}\rightarrow y_2\\
\dots\\
w\xy x_n\rightarrow \colorbox{blue}{\textcolor{white}{\(Control_n\)}}\rightarrow y_n
\end{array}\)} \rightarrow \mbox{Output }\vec{y}
	& = & f(w\xy \vec{x}) =  (y_1,\dots y_n)
\end{array}\]
\caption{Direct product of sequence maps}\label{fig:directc}
\end{figure}

Suppose, on the other hand, that  two controllers, say, \(p\) and \(s\), are interconnected so that the input signal to \(s\) (secondary) should
be the output of controller \(p\) (primary) and the inputs to controller \(p\) should be the pair \((\vec{x}_p,y)\)  where 
\(\vec{x}_p\) is the signal from the ``plant'' and \(y\) is the output from
controller \(s\). Then the connector map can be given:

\[u_i(\ess) = \ess,\quad u_i(w\xy \vec{x}) = \begin{cases} 
u_i(w)\xy Control_{p}(w)&\mbox{if }i=s\\
u_i(w)\xy (\vec{x}_p, Control_{s}(w)) &\mbox{if }i=p\\
u_i(w)\xy \vec{x}_i&\mbox{otherwise}\end{cases}.\]

where \(u_s(w)\xy Control_p(w)\) appends the output of controller \(p\) in the state determined by \(w\) as an event symbol to the sequence
\(u_s(w)\).

\subsection{A system of interconnected processes \label{sec:process}}
Consider a system of connected processes with synchronous message passing.
The specification can be split into a specification of processes \(P:A^*\to X\) for process event alphabet \(A\), followed by a specification of the interconnects.
\begin{figure}[ht]
\[
	\begin{array}{c}\mbox{Inputs: }A\\ \hline \mathit{step}\\ (input,v)\\(\mathit{wrote})\end{array} \rightarrow
\colorbox{blue}{\textcolor{white}{\(P\)}}\rightarrow\begin{array}{l}\mbox{Outputs: }X\\ \hline \mathit{silent}\\ (read, j),\\(write,j,v) \end{array}
\]
	\caption{An abstract process}\label{fig:process}
\end{figure}
Suppose there is a set \(I\) of process identifiers 
and each process has output either \((read, j)\) to request a message from process \(j\), or  \((write, j,v)\) to send value \(v\) to process \(j\), or
\(silent\) to do some internal processing with no I/O in this state. The events then are pairs \((input, v)\) to receive value \(v\),
\(\mathit{wrote}\) for success in sending a message, and \(step\) for some internal computation.  Each of these is a discrete event
symbol, not an expression of any sort (see figure \ref{fig:process}).

Suppose for each \(i\in I\), \(P_i\) is a process. The processes could all be identical or could be different. 
Connection maps \(u_i:B^*\to A^*\) for composite system event alphabet \(B\) determine how the processes are connected.
Then \(P_i(u_i(w))\) is the output of process \(i\) in the state determined by \(u_i(w)\).

\(u_i(\ess)=\ess\) and 
\[u_i(w\xy b) = \begin{cases}
u_i(w)\xy \mathit{step} &\mbox{if } P_i(u_i(w))=\mathit{silent}\\
u_i(w)\xy (input, v) &\mbox{if } P_i(u_i(w))=(read, j)\mbox{ for some }j\\
&\mbox{and } P_j(u_j(w))=(write, i, v)\\
	u_i(w)\xy \mathit{wrote} &\mbox{if } P_i(u_x(w))=(write,j,v)\mbox{ for some }v\\
&\mbox{and } P_j(u_j(w))=(read, i)\\
u_i(w)&\mbox{otherwise}
\end{cases}\]

This specification provides synchronous communication so that a process waiting for a message from process \(j\) or trying to send a message to process \(j\) will make no progress until process \(j\) has
a matching output - if ever. Such systems are known for their propensity to deadlock: e.g. if \(P_i(u_i(w))=(read, j)\) and \(P_j(u_j(w))=(read, i)\) or if larger cycles are created. 
Some solutions require strict ordering of interactions, but it's more common to add timeouts. Suppose we have a map \(\tau:B\to\mathbb{R}\) associating each discrete system event in \(B\) with a duration in seconds. 
Define \(Blocked_i(\ess)=0\) and 
\[Blocked_i(w\xy b) = \begin{cases}
	\tau(b)+Blocked_i(w)&\mbox{if }P_i(u_i(w)) = (write,j,v)\mbox{ for some }j\\
&\mbox{   and }P_j(u_j(w))\neq (read, i)\\
&\mbox{or }P_i(u_i(w))) = (read,j)\\
&\mbox{   and }P_j(u_j(w))\neq (write,i,v)\mbox{ for any }v\\
0&\mbox{otherwise}\end{cases}\]
Then we could add a new \(\mathit{timeout}\) event to \(A\), a constant \(\kappa\) and an additional  case to \(u_i\) where \(u_i(w\xy b) = u_i(w)\xy \mathit{timeout} \) if \(Blocked_i(w) > \kappa\).
By way of contrast, the network in section \ref{sec:big} is completely asynchronous. 

The specification so far is completely parallel and synchronous  - each process advances by one step on each network step if it can. One alternative would be to add a
scheduler so only scheduled tasks advance (see appendix \ref{appendix:sched} for details). 

\section{The Paxos protocol \label{sec:big}}

The first part of this section  is a general packet network connecting network agents. The second part adds constraints to the agents so they obey the Paxos protocol. The goal is to supplement the exposition in ``Paxos Made Simple'' \cite{paxosmadesimple}.

\newcommand{\nullm}{\mathit{NoMessage}}
\newcommand{\xmit}{\mathit{xmit}}
\subsection{Packet network}
\newcommand{\Ids}{\mathit{Ids}}
Consider a networked agent which could be a device, a program, or an
operating system, and only changes state when it receives or sends a message or does some internal
computation.
Each agent that will be connected
on the network has a unique identifier. There is 
a set of messages \(\Msgs\), a set of identifiers \(\Ids\)
and a map 
\(source:\Msgs\to \Ids\) which is intended to tag a message with the identifier
of the agent that sent it. 
\begin{definition}\label{def:G}
	\(G:\Msgs^*\to \Msgs\) is a network agent with identifier
	\(i\in \Ids\) only if:
	 \[\mbox{If }G(w)\neq 0\mbox{ then }source(G(w))=i\]
  (it correctly labels the source of messages
	it sends.)
\end{definition}
The null message \(0\in\Msgs\) is used both for agent output
to indicate the agent has nothing
to send and as an input to indicate the network has nothing to deliver. 
\begin{definition}
\label{def:rxd}  The cumulative set of messages received in the state determined by \(q\in \Msgs^*\) is given by: \[\rxd(\ess)=\emptyset,\quad \rxd(q\xy m) = \begin{cases} \rxd(q)\cup\set{m}&\mbox{if }m \neq 0 \\ \rxd(q)&\mbox{otherwise}\end{cases}\]
\end{definition}
The null message is ignored. 
\begin{definition}

\label{def:txd}
The cumulative  set of messages transmitted by \(G\) in the state determined by \(q\) is given by:
	\[\txd(G, \ess)=\emptyset,\quad\txd(G, q\xy m) = \begin{cases} \txd(G, q)\cup\set{ G(q)}&\mbox{if }G(q)\neq 0\\
	\txd(G,q)&\mbox{otherwise}\end{cases}\] 
\end{definition}
If \(G(q)=0\) then \(\txd(G,q\xy m)\) is unchanged. If \(G\) and \(G'\) are both agents, there
is no requirement that \(\txd(G,w) = \txd(G',w)\) even if both have the same
identifier. 

Note that once a message is in \(\rxd(q)\) or \(\txd(G,q)\) it's there forever - new messages can put a new element
in one of those sets but cannot remove any messages.

\begin{lemma}\label{lemma:netid}
If \(G\) is a network agent with identifier \(i\) and \(m\in \txd(G, q)\) then \(source(m)=i\).
\end{lemma}
The proof is simple, but  the same inductive proof
method is used here multiple times
so it's worth going into a bit of detail.
By the definition of \(\txd\), \(\txd(G,\ess)=\emptyset\) 
so the claim is trivially true for a sequence of length \(0\). Suppose
the claim is true for \(q\). If \(m\in \txd(G,q)\) then \(source(m)\)
doesn't change with any event so the claim is true for \(q\xy a\). If 
\(m\not\in \txd(G,q\xy a)\) the claim is also trivially true. But if 
\(m\not\in\txd(G,q)\) and \(m\in \txd(G,q\xy a)\) then, by the definition of
\(\txd\), \(G(q)=m \neq 0\) which, by definition \ref{def:G} where \(G\) has
identifier \(i\) requires that  \(source(G(q))=i\). 

\begin{definition}
\label{def:net}
A \emph{standard packet network} consists of a collection of 
	 agents, \(G_i: i\in \Ids\) where each \(G_i\)
	has identifier \(i\), plus connection maps
	\[u_i:B^*\to \Msgs^*\]
	where \(B\) is the network event alphabet and
	\begin{enumerate}

		\item Every agent begins in the initial state and each
			system event causes one or zero events for each
			agent:
			\(u_i(\ess)=\ess\) and \(u_i(w\xy a) = u_i(w)\) or
			\(u_i(w\xy a)= u_i(w)\xy m\) for some \(m\in \Msgs\).
\item A non-zero message can only be delivered if it was previously sent:\\
	\(u_i(w\xy b) = u_i(w)\xy m\) for \(m\neq 0\)  only if for some \(j\),\(m\in \txd(G_j,u_j(w))\). 
\end{enumerate}
\end{definition}
\begin{lemma}\label{lemma:rt}
	If \(m\in \rxd(u_i(w))\) then \(m\in \rxd(u_i(w\xy a))\)\\
	and
	if \(m\in \txd(G_i,u_i(w))\) then \(m\in \txd(G_i,u_i(w\xy a))\)
\end{lemma}
The proof follows immediately from the definitions of \(txd\), \(rxd\), and from the standard
\(u_i\) property that \(u_i(w)\) must be a prefix of \(u_i(w\xy b)\).
\begin{lemma}\label{lemma:netsource}
	\(m\in \rxd(u_i(w))\) only if for \(j=source(m)\),\ \(m\in \txd(G_j,u_j(w))\)
\end{lemma}
Proof: As usual, this is trivially true for \(\ess\). Suppose \(m\not\in\rxd(u_i(w))\)
but \(m\in \rxd(u_i(w\xy b))\). Then by the definition of \(\rxd\) and \(u_i\) we know
that \(u_i(w\xy b) = u_i(w)\xy m\) and by definition \ref{def:net}  \(m\in \txd(G_j,u_j(w))\) for some \(j\).
Lemma \ref{lemma:netid} completes the proof.

The network can lose or reorder messages, but can never deliver spurious messages. 
In the sequel, \(q\) will always be used as a variable over \(\Msgs^*\)
and \(w\) over network sequences. Properties of agents that are local, that are true for all
\(q\), do not depend on the network interconnect.

This network is not all that much - in fact, it is almost 
exactly the network that was specified  for a famous impossibility ``theorem'' \cite{Lynch} that says there is no algorithm for detecting whether an agent has failed or is just slow to respond 
from its inputs and outputs. The trick is that, so far, there is no notion of time for agents, so arbitrary pauses are possible and not detectable. 

\subsection{Paxos\label{sec:paxos}}
\newcommand{\notpx}{\mathit{NotPaxos}}
\newcommand{\prepare}{\mathit{Prepare}}
\newcommand{\prepareaccept}{\mathit{PrepareAck}}
\newcommand{\propose}{\mathit{Proposal}}
\newcommand{\proposeaccept}{\mathit{ProposalAck}}
 Paxos\cite{paxosmadesimple} 
 is a 2 phase commit protocol with a twist\cite{Gray}. The protocol
 relies on two sets of agents, a set \(P\) of proposers and a set
 \(C\) of acceptors. There are 4 non-zero message types used in the protocol and
 type \(0\) indicates the message is not part of the protocol:
   \[PTypes=\set{\prepare,\prepareaccept,\propose,\proposeaccept,0}.\]  
 A ``proposer'' agent first requests permission to use a sequence number by sending a ``prepare''
 message that carries a sequence number.
 If and when a quorum of acceptors agree by sending back
 ``prepare accept'' messages with the sequence number,
 the proposer can send a ``proposal''
 with a proposal value and the same sequence number. When or if  a majority of acceptors  agree to the
 proposal by sending proposal accept messages, the proposal has ``won''. 
 The 2 phase twist is that during the prepare accept phase, the proposer can be forced to adopt a proposal value already tried by 
 some lower numbered proposal. This is the most complex part of the protocol
 (see rule \ref{px:priorv} and \ref{px:prior} below) and it produces a result that multiple proposers can ``win'' a consensus, but they must all end up using the same value. 

 \newcommand{\Pmsgs}{\mathit{Pmsgs}}
 \newcommand{\Ptypes}{\mathit{Ptypes}}
    A \emph{Paxos group}   consists of  a standard network, two subsets of ids \(P\) (proposers) and \(C\) (acceptors), a constant \(\kappa>0\) that is more than half of the number of elements of \(C\) (a quorum count),  and a
    set \(\Pmsgs\subset \Msgs\) of messages used in the protocol. There
    is a map \(T:\Msgs\to \Ptypes\) to identify the Paxos message type of
    every message. 
    Each Paxos message has a sequence number  \(seq:\Pmsgs\to \mathbb{N}\). Proposal messages have 
    a value, if \(T(p)=\propose\) then \(val(p)\) is defined.
    Prepare accept messages may carry a proposal message \(p=prev(m)\)
    which is the highest numbered proposal that the acceptor had accepted 
    when it sent the prepare accept. \(Prev(m)= 0\) if the acceptor
    had not accepted any proposals. This will be made precise below.
  Finally, a map \(\pi:\mathbb{N}\to P\) partitions sequence numbers so that  no two proposers are associated with the same
  sequence number. 

  \newcommand{\Accepted}{\mathit{Accepted}}
  \newcommand{\Inherit}{\mathit{Inherit}}
  None of the constants or maps above depend on state but
  the process of controlling proposal values involves two state dependent
  maps that are both ``local'' in the sense that they do not depend on network
  connectivity. \(\Accepted(i,q)\) is the set of proposal messages
  site \(i\) has received and then accepted by transmitting a proposal accept message  with
  a matching sequence number.
  %\[\Accepted(i,\ess)=\emptyset\]
  \[\Accepted(i,q)=\left\{\begin{array}{l}
	  p\in \rxd(q):T(p)=\propose\\
	  \mbox{and for some }m\in \txd(G_i,q), T(m)=\proposeaccept\\
  \mbox{and }seq(m)=seq(p)\end{array}\right\}\]
  \(\Inherit(n,q)\) is the proposal with the highest sequence
  number that has been carried in the \(prev\) attribute
  in a received prepare accept message
  with sequence number \(n\).
  \[\Inherit(n,\ess)=0\]
  \[\Inherit(n,q\xy m)=\begin{cases}
	  prev(m)&\mbox{if }T(m)=\prepareaccept\\
	  &\mbox{ and }seq(m)=n \mbox{ and }prev(m)\neq 0\\
  &\mbox{ and either }(\Inherit(n,q)=0\\
	  &\quad \mbox{ or }seq(\Inherit(n,q))<seq(prev(m)))\\
  \Inherit(n,q)&\mbox{otherwise}\end{cases}\]

The Paxos algorithm can be expressed in 4 rules that control when an agent can transmit a Paxos message
of each non-zero type.
All four rules are local to agents.
 \begin{enumerate}[resume]
	 \item %\textbf{Prepare.}
  \label{px:prepare} \(T(G_i(q))=\prepare\) only if 
  \begin{enumerate}
  \item \(i\in P\)
  \item \label{px:map} and \(seq(G_i(q)) > 0\) and  \(\pi(seq(G_i(q)) = i\)
       \item  and there is no \(m\in \txd(G_i,q)\) with \(T(m)=\prepare\) and  \(seq(m)> seq(G_i(q))\). \end{enumerate}
\item %\textbf{Prepare Accept.}
\label{px:paccept}  \(T(G_i(q))=\prepareaccept\) only if 
\begin{enumerate}
    \item  \(i\in C\)
    \item \label{px:0}and there is some \(m\in \rxd(q)\) with \(T(m)=\prepare\) and \(seq(m)=seq(G_i(q))\)
    \item \label{px:2} and there is no \(m\in \txd(G_i,q)\), where \(T(m)=\prepareaccept\) and \(seq(m) >  seq(G_i(q)).\) 
    \item \label{px:3} and there is no \(m\in \txd(G_i,q)\), where \(T(m)=\proposeaccept\) and \(seq(m) \geq  seq(G_i(q)).\) 
    \item \label{px:prior} and either \(\Accepted(i,q)=\emptyset\)
	    and \(prev(G_i(m))=0)\) or \(prev(G_i(q))\) is the element
		of \(Accepted(i,q)\) with the highest sequence number).
   \end{enumerate}
   \item % proposals
\label{pxrepro} \(T(G_i(q))= \propose\)  only if 
\begin{enumerate}[resume]
\item \(i\in P\)
    \item and \label{px:proposepre} there is some \(m\in \txd(G_i,q))\), \(T(m)=\prepare\) and \(seq(m)>  seq(G_i(m))\)
    \item \label{px:uniqueval} and there is no \(p\in \txd(G_i,q)\) with \(T(p)=\propose\) and \(seq(p)\geq seq(G_i(q))\).
    \item \label{px:procount} and \(\set{source(m): m\in \rxd(q), T(m)=\prepareaccept\mbox{ and } seq(m)= seq(G_i(q))} \) has \(\kappa\) or more elements.
  \item  \label{px:priorv} and if either \(\Inherit(seq(T_i(q)),q)=0\) or
	  \((\Inherit(seq(T_i(q)),q)\neq 0\)\\and \(val(G_i(q))=val(\Inherit(seq(T_i(q)),q))\)
    \end{enumerate}
 \item %\textbf{Proposal Accept. }
 \label{px:accept} \(T(G_i(q))=\proposeaccept\) only if 
 \begin{enumerate}
     \item  \(i\in C\)
     \item \label{px:acceptv} and there is some \(p\in \rxd(q)\) with \(T(p)=\propose, seq(p)=seq(G_i(q))\) 
     \item \label{px:acceptmin} and there is no \(m\in \txd(G_i,q)\) where  \(T(m)\in\set{\proposeaccept,\prepareaccept}\) and \(seq(m)> seq(G_i(q)\).
 \end{enumerate}
\end{enumerate} 

A proposal message \(p\)   ``wins'' in the state determined by sequence \(q\) if the 
proposing agent \(i\) has sent the proposal message and received proposal accept messages
for \(p\) that have source identifiers from \(\kappa\) or more agents. 
This is a property of agents, not the network, because agents need to be able to decide if a proposal has won on the
basis of local data. 
\begin{definition}
\label{definition:win}  \(Wins(i,q,p)\) if and only if
\[\begin{array}{l}
T(p)=\propose\mbox{ and } p\in \txd(G_i,q)\\
	\mbox{and }A(p,q) = \set{source(m): m\in \rxd(q), T(m)=\proposeaccept
\mbox{ and } seq(m)=seq(p) }\\\mbox{ has } \kappa \mbox{ or more elements.}
\end{array}\]
\end{definition}
Suppose \(Wins(i,u_i(w),p)\) is true. 
By lemma \ref{lemma:netsource} each
propose accept message \(m\) received by agent \(i\) must
have been sent by \(source(m)\) so \(\kappa\) or more agents sent propose accept
messages to agent \(i\). By rule \ref{px:accept} each of those agents must
be acceptors (with identifier in the set \(C\)). 

 This is a network property. The event sequence \(w\) determines network state
 and \(u_i(w)\) determines the state of agent \(i\) and depends on \(w\). 

\begin{theorem}\label{th:wins}
If \(Wins(i,u_i(w),p)\) and \(Wins(j, u_j(w),p')\)  then \(val(p)=val(p')\).
\end{theorem}

\subsection{Proof sketch}
See appendix \ref{appendix:paxos} for some details, but the proof can be outlined as follows.
The theorem is proved by creating a list in sequence number order:
\[L(w) = p_0, p_1\dots p_n\]
where \(p_0\) is the winning proposal with the \emph{least } sequence number of any winning proposal and the other elements are all the  
proposals
so that for some \(j\),  \(p\in \txd(G_j,u_j(w))\) and \(seq(p)>seq(p_0)\) .
No proposals with a number less than \(seq(p_0)\) can have won, because \(p_0\) is picked to be the winning proposal with the least sequence number.
Every winning proposal must have been sent by some \(G_j\) from the definition
of \(Wins\). So the list contains
every winning proposal and perhaps some that didn't win but were just proposed.
By rule \ref{px:uniqueval}, no proposer ever sends two different 
proposals with the same number. By rule \ref{px:proposepre} a proposer can only send a proposal if it has previously sent a prepare proposal with the same number and by
rule \ref{px:map}  that number must map to the id of the proposer under \(\pi\) so no two proposer agents can use the same proposal
number. From these considerations the elements of the list never have duplicate numbers and can be strictly sorted by sequence number.

Every proposal \(p_x\) on the list with \(0< x\leq n\) must have had its prepare proposal accepted by \(\kappa\) or more 
acceptors  and \(\kappa\)
or more acceptors sent proposal accept messages for \(p_0\) there must be at least one
acceptor that has sent both kinds of messages. Suppose one of those acceptors has
identifier \(c\):
\[m_0\in \txd(G_c,u_c(w))\mbox{ and } m_x\in \txd(G_c,u_c(w))\]
where \(m_0\) is a proposal accept message with \(seq(m_0)=seq(p_0)\) and 
\(m_x\) is a prepare accept message with \(seq(m_x)=seq(p_x)\). 
By rule \ref{px:3} when \(G_c\) sent \(m_0\) it could not have already sent
\(m_x\). So when \(G_c(q)=p_x\) it must be that \(m_0\in \txd(G_c,q)\). But
to send \(m_0\), \(c\) must have received \(p_0\) so \(\Accepted(c,q)\neq \emptyset\)
and in fact the proposal with the highest sequence number in \(\Accepted(c,q)\)
must have sequence number at least equal to \(seq(p_0)\). That proposal cannot
have sequence number greater than \(seq(m_x)\) by rule \ref{px:3}. In fact, 
the proposer of \(p_x\) can only have received  prepare accept messages
with sequence number less than \(seq(p_x)\). It follows that when that proposer
sent \(p_x\), it was forced to adopt an inherited value from some proposal
with sequence number greater than or equal to the sequence number of \(p_0\) but
less than the sequence number of \(p_x\). Thus, the proposer of \(p_1\) on the
list must have inherited the value of \(p_0\), the proposer of \(p_2\) must
have inherited either the value of \(p_0\) or the value of \(p_1\) which is the same
as the value of \(p_0\) and so on. 

Every winning proposal must be on the list and must
have the same proposal value so 
theorem \ref{th:wins} is proved.

\subsubsection{Discussion}
The specification here is compact but more detailed than the original one in \cite{paxosmadesimple} . For example, agents are defined to output at most one message in each
state --- something that appears to be assumed but  not stated in the original specification. If \(G_i(q)\) could be a set\footnote{For example, if the agent is multi-threaded without appropriate locking.}, an agent could satisfy the specification
and send an accept for \(p_0\) and a prepare accept for \(seq(p)\) at the same time - so that the prepare accept
did not include a previous proposal with a number greater than or equal to 
\(seq(p_0)\). Or consider what happens if an agent transmits a proposal  before any proposal 
has won after receiving \(\kappa\) prepare accepts  
and then receives an additional prepare accept with a higher numbered prior proposal.
The ``Paxos made simple'' specification does not account for this possibility and would
possibly permit the agent to send a second proposal with a different value but the same sequence number but 
rule \ref{px:uniqueval} forbids it here. This is the kind of detail that it's better to nail down in the specification before writing code. 

The proof is intended to convey an intuition about how the protocol works, but to also 
be sufficiently precise considering that it relies on the states of multiple connected
components that change state in parallel.

For contrast, see proofs using formal methods and proof checkers in \cite{paxostla} and \cite{paxisproof2}.

The proof only proves the system is ``safe''  (at most one value can win). Proving liveness (that some value \emph{will} win) is a problem, because neither the 
network as specified (which doesn't need to ever deliver any messages) or Paxos (which can spin issuing new higher numbered proposals that block others, indefinitely) is live as specified.
It would be possible to fix the specification to: rate limit the agents, require eventual delivery from the network, and put timeouts on retries (as done in some implementations \cite{paxoslive} ).

\section{Primitive Recursive Sequence Functions and State Machines\label{sec:math}}

\subsubsection{Standard Representation of Moore Machines}
The standard representation of  a Moore type state machine\cite{Hopcroft} is a sextuple:
\begin{definition}\label{def:moore}
A  \emph{Moore machine tuple} consists of:
\[ M= (A,S,\sigma_0,X, \delta,\lambda)\]
where \(A\) is a set of discrete events (or ``input alphabet''), \(S\) is a state set, \(\sigma_0\in S\) is the start state, \(X\) is a set of outputs
and \(\delta:S\times A\to S\) and \(\lambda:S\to X\) are, respectively, the transition map and the output map. 
\end{definition}
The state set and alphabet are usually required to be finite, but here we sometimes don't need that restriction. 
The next step is to define a class of functions that are equivalent to Moore machine tuples in a strong sense. 

\subsubsection{Characteristic function of a Moore machine tuple}
The set \(A^*\) is the set of finite sequences over the set \(A\), including the empty sequence \(\ess\).
For a finite sequence \(w\) and some \(a\in A\) let \(w\xy a\) be the sequence
obtained by appending \(a\) to \(w\) on the right. 

Each Moore machine tuple is associated with a unique  map \(f:A^* \to X\). 
\begin{definition}\label{def:characteristic}
   The \emph{characteristic sequence map} of Moore machine tuple  \(M= (A,S,\sigma_0,X, \delta,\lambda)\) is :
\[ f_M(w) = \lambda(f'_M(w))\mbox{ where } f'_M(\ess) = \sigma_0,\quad f'_M(w\xy a) = \delta(f'_M(w),a).\] 
\end{definition}

Distinct Moore machine tuples can have the same characteristic sequence map: for example adding non-reachable
states or duplicate states or even changing the state names produces a distinct tuple.  One of the advantages
of working with sequence maps is that these differences are not important in the context of specifying how
systems behave. 

\subsubsection{Sequence primitive recursion}
Sequence primitive recursion \cite{PeterComputer} is a generalization of arithmetic primitive recursion \cite{Peter} so that
 the set \(A^*\)  takes the place of the natural numbers, the 
empty sequence \(\ess\) takes the place
of \(0\) and ``\(+1\)'' corresponds to \(w\xy a\).

\begin{definition}
A map \(f:A^* \to X\) is \emph{sequence primitive recursive} (s.p.r) if and only if
 there is a  set \(Y\), a constant \(c\in Y\), and maps \(h:Y\to X\) and 
\(g:Y\times A\to Y\) so that 
\[f(w) = h(f'(w))\]
where \(f'(\ess)=c\) and  \(f'(w\xy a) = g(f'(w),a)\) for all \(w\in A^*\) 
and \(a\in A\).
\end{definition}
Call, \((c,h,g)\) a \emph{basis} for \(f\) (the set \(Y\) is implicit in the image of \(g\)).

As a very simple example from section \ref{sec:small} consider
\[Counter(\ess) = 0\mbox{ and } Counter(w\xy a) = Counter(w)+1\bmod k\] 
Let   
\(\iota(x)=x\) be the identity map and  \(g(n,a) = n+1\bmod k\) then  \((0,\iota,g)\) is a basis for \(Counter\).
While this may seem trivial, section \ref{sec:composition} shows how
a basis provides a method for simplifying complex sequence maps and showing
they are s.p.r. . 

\begin{lemma}\label{lemma:characteristic} The characteristic sequence map of Moore machine tuple is s.p.r.\end{lemma}
For proof, take the map \(f_M\) from definition \ref{def:characteristic} above and let 
\((\sigma_0,\delta,\lambda)\) be a basis for 
\(f(w) = \lambda(f'(w))\) where \(f'(\ess) = \sigma_0\) and \(f'(w\xy a) = \delta(f'(w),a)\). 
Induction on the length of \(w\) shows \(f = f_M\).

Each s.p.r. function is associated with a unique Moore machine tuple. 

\begin{definition}
If \(f:A^*\to X\) is s.p.r. with basis \((c,g,h)\) so that  \(g: Y\times A\to Y\) and \(c\in Y\),  then 
\[M_f = (A,Y,c, X, g,h)\]
where 
\(X=  \set{h(s): s\in Y}\) is the \((c,g,h)\) \emph{Moore machine tuple}. 
\end{definition}

\begin{lemma}
If \(f\) is s.p.r. with basis \((c,g,h)\)  and \(M_f\) is the \((c,g,h)\) Moore machine tuple, then   the characteristic map of \(M_f\) is the original s.p.r. map \(f\). 
\end{lemma}
The proof is an immediate consequence of the definitions. 

Since each Moore machine tuple is associated with a unique s.p.r. map and each s.p.r. map with a particular
basis is associated with a 
unique Moore machine tuple, s.p.r. maps constitute an alternative representation of the same mathematical objects
that Moore machine tuples represent. 

\subsubsection{Finite state machines}
\begin{definition}
A sequence primitive recursive map \(f:A^* \to X\) is \emph{finite state} only if it has some  s.p.r. basis \(c,h,g\) where \(g\) has a finite image (range).
\end{definition}

\begin{lemma} If \(M\) is finite state then  \(f_M\) is finite state 
and for any  finite state s.p.r. map  \(f\)then for some basis \((c,h,g)\) of  \(f\), the \((c,g,h)\) Moore
machine tuple \(M_f\) is finite state.
\end{lemma}
Proof: if \(M\) is finite state, then \(\delta\) has a finite image, so \(f_M\) is finite state. Conversely, if
\(f\) is finite state with basis \((c,h,g)\) then \(Y\) is a finite set.  

\subsubsection{Via Myhill equivalence}
Instead of using the primitive recursive reduction of a map to a state machine tuple, the connection can be made via the Myhill equivalence \cite{RabinScott}.

For any map: \(f:A^*\to X\), 
let \(w\sim_f q\) if and only if \(f(w\concat z) = f(q\concat z)\) for all \(z\in A^*\). Let \( [w]_f = \set{z: z\sim_f w}\) and 
then put \(S_f = \set{[w]_f: w\in A^*}\). Make \([\ess]_f\) be the start state and \(\delta([w]_f,a) = [w\xy a]_f\) and \(\lambda([w]_f) = f(w)\).

\subsection{Composition \label{sec:composition} } 

Several apparently more complex types of maps can be shown to be sequence primitive recursive. 

%\begin{definition}\label{pr:parallel}
\subsubsection{Direct product}
If  there are  sequence primitive recursive maps \(f_i:A^*\to X_i\) for \(i=1\dots,n\) then let:
\[ f(w) = (f_1(w),\dots f_n(w))\]
This describes a system constructed by connecting multiple components that change state in parallel, without communicating
as shown in figure \ref{fig:directf}.
\begin{figure}[ht]
\[\mbox{input sequence }w \rightarrow 
\begin{array}{lcl}
\fbox{\(\begin{array}{l}
w\rightarrow \colorbox{blue}{\textcolor{white}{\(f_1\)}}\rightarrow x_1 \\
w\rightarrow \colorbox{blue}{\textcolor{white}{\(f_2\)}}\rightarrow x_2\\
\dots\\
w\rightarrow \colorbox{blue}{\textcolor{white}{\(f_n\)}}\rightarrow x_n
\end{array}\)} \rightarrow \mbox{Output }x
& = & w\rightarrow \colorbox{blue}{\textcolor{white}{\(f\)}}\rightarrow (x_1,\dots x_n)
\end{array}\]
\caption{Direct product of sequence maps}\label{fig:directf}
\end{figure}

Claim: \(f\) is sequence primitive recursive. Proof: Let  \((c_i,g_i,h_i)\) be a basis for each \(f_i\) so
that \(f_i(w) = h_i(f'_i(w))\), \(f'_i(\ess)=c_i\) and \(f'_i(w\xy a) = g_i(f'_i(w),a)\) and
\(g_i: Y_i\times A\to Y_i\). 

Let  \(G((y_1,\dots y_n),a) = (g_1(y_1,a),\dots g_n(y_n,a))\) and  \(c=(c_1,\dots c_n)\).
Let   \(H((y_1,\dots y_n)) = (h_1(y_1),\dots h_n(y_n))\).

Then let  \(r(\ess)=c\) and \( r(w\xy a) = G(r(w),a)\).  Clearly \(r\) is s.p.r..
Now we prove \(r(w) = (f'_1(w),\dots f'_n(w))\) by induction on \(w\).
\[\begin{array}{l}
r(\ess) = (c_1,\dots c_n)= (f'_1(\ess),\dots f'_n(\ess))\\
\mbox{Suppose }r(w) = (f'_1(w),\dots f'_n(w))\\
\mbox{then } r(w\xy a) = G(r(w),a) \\
= G((f'_1(w),\dots f'_n(w)),a)\\
= (g_1(f'_1(w),a),\dots g_n(f'_n(w),a))\\
= (f'_1(w\xy a),\dots f'_n(w\xy a))
\end{array}\]

Then \(H(r(w)) = f(w) \) so \(f\) is s.p.r.

%\begin{definition}
\subsubsection{Embedding} See figure \ref{fig:embed}.
\label{pr:embed} If \(f_1\) is s.p.r., let  \(f(\ess) = \kappa\) and \(f(w\xy a) =  g((f(w), f_1(w)), a)\) for 
s.p.r. map \(f_1\).\\
%\end{definition}
	Claim: \(f\) is s.p.r.. Here a new p.r. sequence map is being constructed to depend both on \(w\) and on the values of \(f_1(w)\).
	\begin{figure}
\[ \mbox{Input sequence } w \rightarrow \fbox{ \(w\rightarrow\)  \colorbox{blue}{\textcolor{white}{\(f_1\)}}\(\rightarrow w,x_1 \rightarrow \)\colorbox{gray}{\textcolor{white}{New}}}\rightarrow x.\]
\caption{Embedding a map in a new map}\label{fig:embed}
\end{figure}
Proof. There must be a primitive recursive basis \((c_1, g_1,h_1)\) for \(f_1\) so that \(f_1(w) = h_1(f_1'(w))\) where
\(f'_1(\ess)=c_1\) and \(f'_1(w\xy a) = g_1(f'_1(w),a)\)
\[\begin{array}{l}
\mbox{Let }H((x,y)) = x\\
G((x,y),a) = ( g((x,h_1(y)),a), g_1(y,a))\\
r(\ess)=(\kappa,c_1)\mbox{ and } r(w\xy a) = G(r(w),a)
\end{array}\]
Clearly, \(r\) is s.p.r. Claim: \(r(w)= (f(w),f_1'(w))\). In that case \(H(r(w)) = f(w)\) which proves \(f\) is s.p.r.
\beq
r(\ess) = (\kappa,c_1) = (f(\ess),f_1'(\ess)). \\
\mbox{suppose }r(w) = (f(w),f'_1(w)) \mbox{ then}\\
r(w\xy a) = G(r(w),a) = G((f(w),f'_1(w)),a)\\
= (g((f(w),h_1(f'_1(w))),a),g_1(f'_1(w),a))\\
=  (g((f(w),f_1(w)),a),f'_1(w\xy a))\\
=  (f(w\xy a),f'_1(w\xy a))\\
\mbox{End proof}
\eeq 

\subsubsection{General product} \label{sec:generalf}
\begin{figure}[ht]
\[\mbox{input sequence }w \rightarrow 
\begin{array}{ll}
\fbox{\(\begin{array}{l}
w\rightarrow \colorbox{brown}{\textcolor{white}{\(u_1\)}}\rightarrow \colorbox{blue}{\textcolor{white}{\(f_1\)}}\longrightarrow x_1 \\
w\rightarrow \colorbox{brown}{\textcolor{white}{\(u_2\)}}\rightarrow\colorbox{blue}{\textcolor{white}{\(f_2\)}}\longrightarrow x_2\\
\dots\\
w\rightarrow \colorbox{brown}{\textcolor{white}{\(u_n\)}}\rightarrow \colorbox{blue}{\textcolor{white}{\(f_n\)}}\longrightarrow x_n\\
\quad \gamma\Uparrow\leftarrow (x_1,\dots x_n) \Leftarrow \mbox{ feedback }\Downarrow
\end{array}\)} \rightarrow \mbox{Output }(x_1,\dots x_n) 
&  \rightarrow 
\end{array}\]
\caption{Event and output flow in the general product \label{fig:gproductf}}
\end{figure}

\begin{definition}\label{def:general}
For \(f_i:A_i^*\to X_i\) and \(\gamma_i: X_1\times\dots X_n\times  A\to A_i^*\) where 
\((0 < i \leq n)\) and each \(f_i\) is s.p.r., 
the \emph{general product} is:
	\[\begin{array}{l} f(w) = (f_1(u_1(w)),\dots f_n(u_n(w)))\\
	u_i(\ess)=\ess,\mbox{ and } u_i(w\xy a) =  u_i(w)\concat \gamma_i(f(w),a)\mbox{ for }i=1,\dots n\end{array}\]
where \(\concat\) is the usual concatenation of finite sequences. 
\end{definition}
As illustrated in figure \ref{fig:gproductf}, if we have \(n\) p.r. sequence maps \(f_1,\dots f_n\) with each \(f_i:A i^*\to X_i\), the system output when they are connected is in the set \(X=X_1\dots \times X_n\), and
a ``connector'' is a map \(\gamma_i:A\times X\to A_i^*\) so that \(\gamma_i(a,x)\) is the sequence of events produced for component \(i\) when the system input is \(a\) and the 
outputs of all the components are given by \(x\). The event alphabets of the components can all be different or the same and the composite alphabet can also be different or the same 
depending only on  \(\gamma_i\).

\begin{theorem} If \(f_1,\dots f_n\) are s.p.r. in a product of the type of definition \ref{def:general}, then \(f\) is s.p.r. \end{theorem}
Proof: 
Each \(f_i\) has a basis \((c_i,g_i,h_i) \) with  \(g_i: Y_i\times A_i\to Y_i\) and \(f_i(q)= h_i(f'_i(q))\) where 
\[f'_i:A_i^*\to Y_i \mbox{ and } f'_i(\ess)=c_i\mbox{  and }f'_i(w\xy a) = g_i(f'_i(w),a)\]

Let \(H(y_1,\dots y_n)= (h_1(y_1),\dots h_n(y_n))\) so \(f(w) = H(f'_1(u_1(w)),\dots f'_n(u_n(w)))\). 
The goal is to define   a s.p.r. map \[r:A^*\to Y_1\times \dots Y_n\]
so that \(r(w) = (f'_1(u_1(w)),\dots f'_n(u_n(w)))\), 
which implies \(H(r(w))= f(w)\). This will prove \(f\) is s.p.r..

Because \(\gamma_i\) is sequence valued it is useful to extend each \(g_i\) to sequences:
\[g_i^*:Y_i\times A^*.\mbox{ Let }  g_i^*(y,\ess)=y\mbox{ and }g_i^*(y,q\xy a)= g_i(g_i^*(y,q),a).\]
Then let: 
\[F'_i(\ess)=c_i\mbox{ and }F_i'(q\xy a) = g_i^*(F'_i(q),\ess\xy a).\]
Clearly \(F'_i(q)=f'_i(q)\) so \(f_i(q) = h_i(F'_i(q))\).   
\beq
\mbox{for }y=(y_1,\dots y_n)\mbox{ let } G(y,a) = (g_1^*(y_1,\gamma_1(H(y),\ess\xy a)),\dots g_n^*(y_n,\gamma_n(H(y),\ess\xy a)))\\
\mbox{Let } r(\ess) = (c_1,\dots c_n), \mbox{ and } r(w\xy a) = G(r(w), a)\\
\mbox{By construction }r\mbox{ is s.p.r.}\\
\mbox{Claim }r(w) = (F'_1(u_1(w)),\dots F'_n(u_n(w)))\\
\mbox{Proof by induction on }w\\
r(\ess) = (c_1,\dots c_n)\\
= (F'_1(\ess),\dots F'_n(\ess)\\
= (F'_1(u_1(\ess)),\dots F'_n(u_n(\ess))\\
\mbox{Inductive hypothesis }r(w) = y =  (F'_1(u_1(w)),\dots F'_n(u_n(w)))\\
r(w\xy a) = G(r(w),a)\\
= (g_1^*(y_1,\gamma_1(H(y),\ess\xy a)),\dots g_n^*(y_n,\gamma_n(H(y),\ess\xy a)))\\
= (g_1^*(F'_1(u_1(w),\gamma_1(f(w),a)),\dots g_n^*(F'_n(w),\gamma_n(f(w),a)))\\
= (F'_1(u_1(w\xy a)),\dots F'_n(w\xy a))))\\
QED
\eeq
The proof here is not complicated, but I originally produced it by going via the
Moore machine tuple representation covered in section \ref{sec:moore} as it was easier to visualize. In that proof, first each component map is converted to a Moore machine tuple  and then they are multiplied out in the general product and then the result is converted back to the characteristic s.p.r. map.

 \subsection{Moore machine  products}\label{sec:moore}
The general product of Moore machines \cite{Hartmanis}  and later \cite{Gecseg, Yoipl} has a state set constructed as the cross product of the state sets of the 
factor machine and has a connector map for each component \(\phi_i:X_1\dots \times X_n\times A\to A_i\).
Compare to definition \ref{def:general}. 

\begin{definition}\label{def:mgeneral}
Given a set \(A\) of product machine events,  Moore machine tuples 
\(M_i = (A_i,S_i,\sigma_{i,0}, \delta_i,X_i,\lambda_i)\)  and
connectors 
\(\phi_i:X\times A\to A_i\) for \(i=1,\dots.n\), \emph{the general product:}
\[M= (A,S,\sigma_0,\delta,X,\lambda) \mbox{ where}\]
\(X=X_1\times \dots X_n\), and  \(S = S_1\times \dots S_n\)\\
the initial state \(\sigma_0 = (\sigma_{1,0},\dots \sigma_{n,0})\) \\
the output map \(\lambda((s_1,\dots s_n)) = (\lambda_1(s_1),\dots \lambda_n(s_n))\)\\
and the transition map 
\(\delta((\sigma_1,\dots \sigma_n),a) = (\delta_1(\sigma_1,a_1), \dots \delta_n(\sigma_n,a_n))\) where
\(a_i = \phi_i(\lambda_1(\sigma_1),\dots\lambda_n(\sigma_n)),a)\)

\end{definition}
If each \(M_i\) is finite state, the product is finite state. 

The connectors can be extended to produce sequences on each step just as with the general product of sequence primitive recursive functions. 

Since \(M\) is a Moore type state machine tuple, it has a characteristic map.  If \(f_i\) is a characteristic map for each \(M_i\) then
\[f(w) = (f_1(u_1(w)),\dots f_n(u_n(w)))\]
where each \(u_i(\ess) = \ess\) and \(u_i(w\xy a) = u_i(w)\xy \phi_i(f_1(u_1(w)),\dots f_n(u_n(w)),a))\)
is the characteristic map for \(M\). 

\subsection{Algebra \label{sec:Monoids}}
One of the motivations in this work for staying within the bounds of classical
deterministic state machines is that they have
been shown to be fundamental mathematical
objects directly connected to semigroup and group theory\cite{Pin,Holcombe,Ginzburg,baumslag}.
Each state machine defines a semigroup via  congruence classes on finite sequences as shown by Nerode and Myhill (p. 70-72 \cite{RabinScott}).
Earlier work in algebraic automatic theory looked
at connections between the component structure of computer systems
and the ``loop-free'' product structure of the characteristic monoids.
In the construction of general product of s.p.r. maps, if
each \(\gamma_i\) depends only on the last argument, then the product reduces to a ``direct'' or ``cross'' product and
the state machines are not interconnected.  If each \(\gamma_i(x_1,\dots x_n,a)\) depends only on  \(a\) and 
\(x_1,\dots x_{i-1}\) then the product reduces to a  ``loop free'' or ``cascade'' product \cite{HartmanisStearns,Holcombe,Pin,OMaler}
in which information only flows in a linearly ordered pipeline
through the factors. 
But
the network in  section \ref{sec:big} is an easy example of a system where a loop-free decomposition will not reflect system architecture.

\subsubsection{Modularity\label{sec:modular}}
Moore 
machines distinguish between internal state and
externally visible state (output) in a way that 
corresponds to ``information hiding'' \cite{ParnasCriteria}. This is why the connectors of the general product \ref{def:general} depend on the outputs of the 
components, not on their interior state sets. The product structure
of systems can then provide an insight into modularity
in terms of how much of the internal state of components
must be communicated to other components.  It is relatively simple to show that any finite state map with \(n\) states can be constructed from \(\log_2 n\) single bit state maps - but at the expense of making all state information 
visible and communicating the state of every component to every other component on each step. The ratio of the size of the output set to the size of the set
\(A^*/\sim_f\) indicates the extent of information hiding. That is, the benefits
of modularity cannot be automatically obtained by breaking a system into
components. This is a well known engineering principle \cite{amdahl} which, 
perhaps, can be investigated more in terms of automata products and monoid
structure. For example, the design basis of 
micro-kernel operating systems is appealing, but there might be fundamental
mathematical reasons why obtaining high performance is so difficult\cite{chen}.

\section{\label{sec:related} Related work}
\subsection{Sources}
 Deterministic state  machines\cite{RabinScott}, particularly finite state machines, are well known and 
widely used in computer science and digital circuit engineering.
Moore type machines \cite{Moore}
add a model of interaction as input/output and  of modularity via the distinction between internal state
and visible state (output). The \emph{concurrent product} given by Hartmanis in 1964\cite{Hartmanis} and 
later in \cite{Gecseg} provides semantics for composition, concurrency, and encapsulation.  Algebraic automata theory \cite{Arbib,Holcombe,Ginzburg} includes a view of automata  
as maps from finite sequences of events to output. These 
sequence maps are used in algebraic automata theory to show equivalence between machines that ``do the same thing" even
though they might have different state sets. Here sequence maps are used \emph{in place}  of methods like state diagrams so that state sets do not need to be made explicit and can be parameterized and so that the emphasis can be on behavior. 
Primitive recursive functions on words, are adapted from Rozsa Peter's ``Recursive functions in Computer Theory"  \cite{PeterComputer} and``Recursive Functions'' \cite{Peter}. There is a more abstract treatment of the same functions
in \cite{EilenbergElgot}.  Application of primitive recursion to describe state machines was introduced by this author
as a semantics for an extended temporal logic\cite{Yoipl,Yoams,Yocav}. As discussed in the next section, however,
this paper does not
use the methods of formal logics. 
The treatment of interconnection by dependent variables is new to this paper as is the notion of a basis for 
sequence primitive recursive functions.

\subsection{Comparison}

\begin{quote}\textit{
    Everybody who has worked in formal logic will confirm that it is one of the technically most refractory parts of mathematics --- John von Neumann\cite{vonNeumann}}
\end{quote}
\begin{quote}\textit{
engineers in practice are not familiar with and not fond of large
	logical formulas that arise if an untuned logical formalism is used \cite{BroyTUM}}
\end{quote}
This project began with temporal logic
\cite{Manna,Pnueli,Clarke, LamportTLA,Krithi} 
which adds a variety of qualifiers to a first order logic in order
to be able to express \emph{when} propositions can or must become true.
The current version, however, does not use any formalism in the sense of 
formal logics and axiomatic methods but is based on
recursive functions and ordinary applied mathematics.  
The goal is to ``formalize'' system specifications in the sense of expressing 
them precisely and mathematically but not to ``formalize'' in the sense of
syntactic methods employed in the 
foundations of mathematics\cite{schoenfield}
or in the temporal logics or similar.

In TLA\cite{LamportTLA} and related approaches
\cite{AbadiLtoplas}, 
the ``semantics'' consists of infinite sequences of assignment maps, each map
assigns values to variable symbols. A component is specified with a formal
expression in the logic
and is evaluated against the sequence of maps.  Components are
composed by conjunction of their specifications. Because the semantics
doesn't directly support parallel state changes or composition, interleaving
is imposed  axiomatically. The specification of the 
composed system in temporal logic includes clauses that require
that the variables of the specification of
a component system can always be kept constant over the ``next'' state transition so that
some other component advances in that step.
But because this would allow components to never make progress, further clauses
are required to impose ``fairness'' on the sequences. 
S.p.r. maps can express interleaving, scheduling, and parallel state change
directly in terms of products as shown in the examples above.  See appendix \ref{appendix:tl} for more on temporal logics and
interleaving.

FOCUS \cite{BroyTUM,Broy} on the other hand, 
takes named typed channels for ``asynchronous, buffered
message exchange'' as the \emph{primitive} method of communication between components. 
FOCUS semantics is given as maps from and to infinite sequences of messages (or infinite sequences
of finite sequences of messages) labeled with channel identifiers. Systems are specified by 
interface (lists of channels and channel types) and 
formal expressions written in second order logics. The
formal variables are assigned values by the message sequences. 
FOCUS can describe  parallel composition with feedback,
but, as with TLA,  composition is described by the 
``and'' of the specifications on the components, in this case
with some additional clauses 
for shared channel names which determine the component communication.

FOCUS maps are both elaborate  and highly specific about communication,
and at the same time quite general objects --- far more general than 
state machines.  For example, these maps
do not necessarily have 
causal relationships between input and output. That is, it is 
possible for a specification to  define a component that can
respond to events that have not happened yet but that are
further down the infinite sequence of input messages. Since that's not 
a sensible property of a computer system, the formal 
specifications of a system component can be 
be extended to impose ``strong causality'' 
axiomatically.  Broy points out that a strongly causal FOCUS map   ``essentially
defines a deterministic automaton, called Mealy machine'' (\cite{Broy} p.6).
In fact, in hopes of making FOCUS specifications more intuitive\footnote{
``often properties of systems have to be written by too lengthy, unintelligible, intricate formulas.''(\cite{BroyTUM} p.3)}
, they can be combined with state diagrams, where formal predicates
are attached to transitions and states can be arbitrarily complex.

There are similarities between most methods for computer system specification because the final objects being 
specified are the same. S.p.r. maps differ from both TLA and FOCUS because they are constructive not axiomatic or
extensional.
The system design expressed in s.p.r. maps determines interconnection,
concurrency and encapsulation, 
and s.p.r. maps are causal by construction.
FOCUS and TLA (and many similar methods)
require use of formal logics because the underlying semantics is more
general than state machines (and less structured than s.p.r. maps or Moore machines in product form).
System specifications are, consequently, 
more complicated and require something like a formal logic to 
add structure to the semantics. 
Whether this is important to the system designer or not is 
something that has to be determined by experience.

\subsection{Future work}
In addition to the avenues for future research 
noted in section \ref{sec:Monoids} both specification of more real-world
systems,
and some automation would be useful. For the first, more medium sized
examples would produce empirical information on best ways of
carrying out specifications and
proofs and larger examples (operating systems) would test the limits of the 
method. For the second, anyone reading earlier versions of this paper
should immediately see the advantage of programs checking for typographical
errors, type agreement (e.g. ``what is \(p_x\), shouldn't it be \(p_i\)?'') and
even simple dimensional analysis \cite{street}. 
Another useful
automation would analyze specification text against a proof text to see which
assertions have been cited in the proof. More ambitiously,
formal logic is not required for automation of
mathematics, e.g.  see \cite{computeralgebra} , and integrating 
s.p.r. maps and proofs into a computer algebra system would be interesting.
Finally, some visual method, such as Statecharts \cite{HarelState} seems like
it might be practical,  using s.p.r. maps to resolve or evade the 
``extremely delicate problems'' and subtle issues \cite{HarelSemantics} of
the \emph{ad hoc} semantics.

\section{Acknowledgments}
Thanks very much to the reviewers and particularly to the reviewing editor, Ewen Denney, for insights, comments, encouragement
and suggestions. The reviewers careful analysis and close reading was enormously helpful and much appreciated.  
\bibliographystyle{alpha}
\bibliography{sm}
\appendix

\section{Appendix: Some more small examples}\label{appendix:examples}
\subsection{Memory}\label{appendix:memory}
Using the cell defined above in section  define a memory array as in figure \ref{fig:memory}.
\begin{figure}[ht]
\[\begin{array}{rcccl} w \rightarrow&\colorbox{blue}{\textcolor{white}{\(\begin{array}{|c|c|}\hline
\mathit{Cell}_1 & \mathit{Cell}_2 \\  \hline
\mathit{Cell}_3 &\mathit{Cell}_4 \\ \hline 
\mathit{Cell}_6 &\mathit{Cell}_5 \\ \hline 
\end{array}\)}}&\rightarrow 
\colorbox{blue}{\textcolor{white}{Mem}}&\rightarrow 
 \end{array}\]
 \caption{Memory array}
 \label{fig:memory}
\end{figure}
For some set \(D\) of addresses and \(V\) of values, the memory array alphabet \(A\) consists of events: 
\((read,d)\) to read from memory cell \(d\), \((write,d,x)\) to write value \(x\) to cell \(d\), and
\((copy,d,d')\) to copy the contents of cell \(d\) to cell \(d'\). 
All we need is one memory cell map, \(Cell:V^*\to V\) with different input sequences
\(u_d\) for each \(d\in D\).
\[ u_d(\ess)=\ess,\quad u_d(w\xy a) = \begin{cases}
    u_d(w)\xy x&\mbox{if }a=(write,d,x), x\in V\\
	u_d(w)\xy Cell(u_{d'}(w))&\mbox{if }a=(copy,d',d), d'\in D\\
    u_d(w)&\mbox{otherwise}
\end{cases}\]
The connection maps follow a standard scheme \(u_d(\ess)=\ess\) so the components all start in their
initial states, and either \(u_d(w\xy a) = u_d(w)\xy v\) for some cell value \(v\)
or \(u_d(w\xy a) = u_d(w)\) to leave the cell state unchanged
\footnote{ Where useful, the components can be defined to advance by multiple
steps on a single event of the enclosing system. See section \ref{sec:generalf}}. In this example, the most complicated   
case \(u_d(w)\xy Cell(u_{d'}(w))\) appends the output of cell \(d'\)
to the sequence of events being constructed for cell \(d\).

Define: 
\[Mem(w\xy a) = \begin{cases} 
Cell(u_d(w))&\mbox{if }a=(read,d)\mbox{for some }d\in D\\
Mem(w)&\mbox{otherwise}\end{cases}
\]
to complete the system.  To extend this specification to support events \((swap,d,d')\) that swap the contents of two cells, change
the definition of \(u_d\) so
\[ u_d(\ess)=\ess,\quad u_d(w\xy a) = \begin{cases}
    u_d(w)\xy x&\mbox{if }a=(write,d,x), x\in V\\
	u_d(w)\xy Cell(u_{d'}(w))&\mbox{if }a=(copy,{d'},d), {d'}\in D\\
    &\mbox{or }a=(swap,d ,d')\mbox{ or }a=(swap,d',d), d,d'\in D\\
    u_d(w)&\mbox{otherwise}
\end{cases}\]
This is a very elementary example, but it's a definition of a deterministic state machine as a product of state machines
where each component can receive input from the external system or from any other cell and cells can change state in parallel with 
each other and the \(Mem\) component.

\subsection{Scheduler}\label{appendix:sched}

There are many possible ways to add a scheduler to the system of processes of section \ref{sec:process}.
Perhaps on each step of the composite system the scheduler picks a set of processes to advance by one step. The inputs to the scheduler on 
each step might be the tuple of outputs of the process components so the input alphabet of the scheduler is \(C = X^n\) where \(n\) is the number of 
elements in \(I\) and the outputs of the scheduler are sets of process ids to advance so a scheduler is \(S:C^*\to J\) where \(J\) contains
some set of subsets of \(I\). Define 
\[v(\ess)=\ess, v(w\xy b) = v(w)\xy (x_1,\dots x_n)\mbox{ where }x_i = P_i(u_i(w))\]

Then modify the other connectors 
\(u_i(\ess)=\ess\) and 
\[u_i(w\xy a) = \begin{cases}
	u_i(w)\xy \mathit{step} &\mbox{if }i\in S(v(w))  \mbox{ and } P_i(u_i(w))=\mathit{silent}\\
u_i(w)\xy (input, v) &\mbox{if } P_i(u_i(w))=(read, j)\mbox{ for some }j)\\
	&\mbox{and } i\in S(v(w))\mbox{ and }j\in S(v(w))\\
&\mbox{and } P_j(u_j(w))=(write, i, v)\\
	u_x(w)\xy \mathit{wrote} &\mbox{if } P_i(u_x(w)=(write,j,v)\mbox{ for some }v\\
	&\mbox{and } i\in S(v(w))\mbox{ and }j\in S(v(w))\\
&\mbox{and } P_j(u_j(w))=(read, i)\\
u_x(w)&\mbox{otherwise}
\end{cases}\]

\section{Appendix: Some Details of the Paxos safety proof}\label{appendix:paxos}

\begin{lemma}\label{lemma:B}
	If an acceptor has sent both a prepare accept message and a propose accept message where
	the proposal accept message has a smaller sequence number, then the prepare accept message must
	carry a prior proposal with a sequence number at least as great as that of the proposal accept.

If \(m_x\in \txd(G_i,q)\) and \(m_y\in \txd(G_i,q)\) and \(T(m_x)=\proposeaccept\) and \(T(m_y)=\prepareaccept\)  and \(seq(m_x) < seq(m_y)\) \textbf{then}  \(seq(prev(m_y)) \geq seq(m_x)\).
\end{lemma}
Proof by induction on sequence length. The lemma is trivially true for length \(0\) since \(\txd(G_i,\ess)=\emptyset\) by definition \ref{def:txd}.
Assume the lemma is true for sequence \(z\) and consider \(z\xy a\). There are 4 cases for each \(z\) and \(a\):
\begin{enumerate}
    \item \(m_y\in \txd(G_i,z)\) and \(m_y\in \txd(G_i,z)\) in which case, by the induction hypothesis 
 \(seq(m_x) \leq seq(prev(m_y)))\) and these are not state dependent so the inequality holds in the state determined by \(z\xy a\).
 \item \(m_x \not\in \txd(G_i,z\xy a)\) or \(m_y\not\in \txd(G_i,z\xy a)\) in which case the lemma is trivially true in the state determined by \(z\xy a\)
 \item  \(m_x \not\in \txd(G_i,z)\) or \(m_y\not\in \txd(G_i,z)\)  and \(m_x \in \txd(G_i,z\xy a)\) or \(m_y\in \txd(G_i,z\xy a)\).
 By the definition of \(\txd\) one of \(m\in \txd(G_i,z)\) or \(m_y\in \txd(G_i,z)\) because at most one element is added to \(\txd(G_i,z)\) by the event \(a\). So there
 are two subcases:
\begin{enumerate}
\item If \(m_y\in \txd(G_i,G_i,z)\) and \(G_i(z)=m_x\) by rule \ref{px:acceptmin} \(T(G_i(z))\neq \proposeaccept\) contradicting the hypothesis. This case cannot happen.  
\item If \(m\in \txd(G_i,G_i,z)\) and \(G_i(z)=m_y\) then, by lemma \ref{lemma:A} there is some \(p\in \rxd(q)\) so \(T(p)=\propose\) and \(seq(p)=seq(m_x)\). The set 
	\(\Accepted(i,q)\) in rule \ref{px:prior} then contains at least \(p\), so \(seq(prev(m_y)) \geq seq(p)=seq(m_x) \).
\end{enumerate}
\end{enumerate}

\begin{lemma}\label{lemma:minq}
	If \(Wins(i,u_i(w),p)\) and \(G_j(u_j(w))=p'\) for some \(p'\) where \(T(p')=\propose\)
	and \(seq(p')> seq(p)\)  then there is a \(m\in \rxd(u_j(w))\), \(T(m)=\prepareaccept\) and \(seq(m)=seq(p'\)
	and \(seq(prev(p'))\geq seq(p)\)
\end{lemma}
This is a ridiculously long chain of assertions, but almost all the work is done. From the definition of \(Wins\), site \(i\) has received proposal accept messages for \(p\) from \(\kappa\) or more acceptors. From 
rule \ref{px:procount} agent \(j\) must have received prepare accept messages for \(seq(p')\) from \(\kappa\) or more acceptors.  Since \(\kappa\) is more than half, there must be at least one agent \(x\) that
has transmitted both a proposal
accept for \(p\) and a  prepare accept \(m\) for \(seq(p')\) from \(x\). By  lemma \ref{lemma:B} then \(seq(prev(m)) \geq seq(p)\). 
\begin{lemma}
	If   \(m\in\txd(G_i,q)\) and (\(T(m)=\prepareaccept\) or \(T(m)=\proposeaccept)\)) then \(i\in C\).\\ 
	This follows directly from rules \ref{px:paccept} and \ref{px:accept} and from  lemma  \ref{lemma:netid}.
\end{lemma}
\begin{lemma}
 If   \(p\in \txd(G_j,q)\) and \(T(p)=\propose\) then \(\pi(seq(p))=j\).
\end{lemma}
Proof:  \(G(q)=p\) implies, by rule \ref{px:proposepre} that there is some \(m\in \txd(G_j,q)\) where \(seq(m)=seq(p)\) and \(T(m)=\prepare\). And rule 
\ref{px:map} requires that \(G_j(z)=m\) only if \(\pi(seq(m))=source(m)\) and lemma \ref{lemma:netid} requires that \(source(m)=j\). Since \(seq(m)=seq(p)\) then
\(source(p)=source(m) = j = \pi(source(p))\).
\begin{lemma}
    If \(p\in \txd(G_j,q)\) where \(T(p)=\propose\)  and \(p'\in \txd(G_j,q)\) where \(T(p')=\propose\) and \(seq(p)=seq(p')\) then \(p=p'\).
\end{lemma}
Proof:    Suppose, without loss of 
    generality that \(p\in \txd(G_j,z)\) and \(G_j(z)=p'\) then by rule \ref{px:uniqueval} \(p= p'\).

It follows that
\begin{lemma}
	If \(p\in \txd(G_j,u_j(w))\) where \(T(p)=\propose\)  and \(p'\in \txd(G_i,s_i(w))\)  then \(j=i\) or \(seq(p)\neq seq(p')\)
\end{lemma}
If \(seq(p)=seq(p')\) then \(\pi(seq(p))= j = \pi(seq(p')) = i \). 

\begin{lemma}\label{lemma:A}
If a agent has sent a proposal accept, it has received a matching proposal (with the same sequence number)\\
If \(m\in \txd(G_j,q), T(m)=\proposeaccept\) then there is some \(p, T(p)=\propose, seq(p)=seq(m), p\in \rxd(q)\).
\end{lemma}
Proof by induction on prefixes of \(q\),  Initially \(m\not\in \txd(G_j,,\ess\). Suppose \(m\in \txd(G_j,,z\xy a\))  but \(m\not\in\ \txd(G_j,,z)\) then
\(G(z)=m\) (by definition of \(txd\))  which means by \ref{px:acceptv} the matching proposal must have been received.
\begin{lemma}\label{lemma:y}
If  \(m\in \txd(G_i,q)\) and \(T(m)=\prepareaccept\) then either \(source(prior(m))=0\) or \(seq(prior(m)) < seq(m))\).
\end{lemma}
Suppose \(G(q)=m\) and \(T(m)=\prepareaccept\). By rule \ref{px:prior} if \(prior(m)\neq 0\) then \(prior(m)=p\) so that
\[ p\in \set{ p\in \rxd(q), T(p)=\propose \mbox{ and }\exists m_c\in\txd(G_i,q), T(m_c)=\proposeaccept, seq(m_c)=seq(p)}\]
So \(m_c\in \txd(G_i,q)\) which implies, by rule \ref{px:3} that \(seq(m_c)< seq(m)\).

\section{Notes on Temporal Logic\label{appendix:tl}}

The kinds of properties that one might express in temporal logics can
also be made precise using s.p.r. maps. For example, the expression
\(P(w)\) where \(P\) is a boolean s.p.r. map and \(w\) is a free variable over
event sequences means ``\(P\) is always true''. 
Define \(NotP(w\xy a) = (1 - P(w))*(1+ NotP(w))\) and then \(NotP(w) < t\) or
``for some \(t\), \( NotP(w) < t\)'' are concrete versions of ``eventually \(P\)''.  
To assert that \(P\) must become true before \(Q\) becomes true, define
\(EdgeP(\ess)=0\), \(EdgeP(w\xy a) = max(P(w\xy a),EdgeP(w))\)
then the desired property is \(Q(w)\leq EdgeP(w)\). 
The proofs in section \ref{sec:big}  involve showing
that some condition must be true
\emph{before} a particular message can be transmitted.
If we wanted to assert message delivery is ``fair'', 
we could count how many times any message has been transmitted and then 
add a probability measure\footnote{One way to do that is to suppose that the
network inputs include some signal that determines the rate 
and distribution of message delivery failure.}. 

The semantics of temporal logics come from a form of  state machines where states are maps from formal variables to
values
\footnote{TLA uses sequences of states, but these can be considered traces through a state machine.}.
The state machines
or sequences are unlabeled and non-deterministic so much of behavior of the system are expressed as 
axioms  in the logical language.
Without automata products, temporal logics model composition and concurrency via
interleaving. 
Using interleaving to model concurrency 
treats a concept from programming languages with concurrent threads as fundamental. Specification of
an operating system, where some operations are truly parallel and some are scheduled by the OS itself does not fit
into this seamlessly.
In this regard see \cite{LamportTLA} p.44,
\begin{quote}
``TLA is based on an interleaving model of concurrency, in which we assume that an execution of the system consists of a sequence
of atomic events. It seems paradoxical to represent concurrent systems with a formalism in which events are never concurrent. We will not attempt to justify the
philosophical correctness of interleaving models for reasoning about concurrent algorithms.''. 
\end{quote}
The semantics is not causal (transitions are unlabeled and considered to represent passage of a unit of time or a step).
You can't say, \(x(w\xy a) = 1\) and \(x(w\xy b)=0\), you have to 
say something like \(next (lastevent=a \rightarrow x=1) \) etc. and then provide additional rules for ``lastevent'' which can easily become quite complex. Since composition involves combining state assignment maps, any rules
for ``lastevent'' must take into account that the other component might have different last events.
or no events at all. There is no semantic connection  between the assignment maps and the state nodes in the graphs --- this has to be specified axiomatically. Additionally, the composed state machines are ``flat'' with execution interleaved non-deterministically. 
To express something similar to \(m\in \rxd(u_i(w)) \rightarrow \rxd(u_j(w))\) would 
involve significant scaffolding.

\end{document}